\documentclass[journal]{IEEEtran}

\ifCLASSINFOpdf
   \usepackage[pdftex]{graphicx}
   \graphicspath{{../pdf/}{../jpeg/}}
   \DeclareGraphicsExtensions{.pdf,.jpeg,.png}
\else
   \usepackage[dvips]{graphicx}
   \graphicspath{{../eps/}}
\fi

\usepackage{threeparttable}
\usepackage{bm}
\usepackage{amsfonts}
\usepackage{booktabs}
\usepackage{multirow}
\usepackage{subfigure}
\usepackage{tabularx}
\usepackage{algorithm}
\usepackage{amsmath}
\usepackage{amssymb}
\usepackage{color}
\usepackage{cite}
\usepackage{algorithmicx}
\usepackage{algpseudocode}
\usepackage{lineno,hyperref}
\usepackage{graphicx}
\usepackage{epsfig}
\usepackage{array}

\usepackage{graphicx}
\usepackage{epsfig}
\usepackage{graphicx}
\usepackage{amsmath}
\usepackage{amssymb}
\usepackage{multirow}
\usepackage{changepage}
\usepackage{subfigure}

\newcommand{\PreserveBackslash}[1]{\let\temp=\\#1\let\\=\temp}
\newcolumntype{C}[1]{>{\PreserveBackslash\centering}p{#1}}
\newtheorem{remark}{Remark}
\newtheorem{theorem}{Theorem}

\newtheorem{proposition}{Proposition}

\begin{document}
\title{Low Complexity Successive Cancellation Decoding of Polar Codes based on Pruning Strategy in Deletion Error Channels}
\author{He Sun, \emph{Member, IEEE,}
        Bin Dai, \emph{Member, IEEE,}
        Rongke Liu, \emph{Senior Member, IEEE}

\thanks{This paper is uploaded as a record and reference for our previous work on deletion channel. It was finished and submitted to TVT in 2020 and has been partly published in ``Reduced-complexity successive-cancellation decoding for polar codes on channels with insertions and deletions'', IEEE Transactions on Communications, 2022.}
\thanks{He Sun and Rongke Liu are with the School of Electronic and Information Engineering, Beihang University, Beijing 100191, China (e-mail: {sunhe1710@buaa.edu.cn, rongke\_liu@buaa.edu.cn}). Bin Dai is with the School of Internet of Things, Nanjing University of Posts and Telecommunications, Nanjing,  210023 China (e-mail: daibin@njupt.edu.cn).}
}
\markboth{ }%
{Shell \MakeLowercase{\textit{et al.}}: Bare Demo of IEEEtran.cls for IEEE Journals}
\maketitle

\begin{abstract}
A novel SC decoding method of polar codes is proposed in $d-$deletion channels, where a new pruning strategy is designed to reduce decoding complexity. Considering the difference of the scenario weight distributions, pruning thresholds for each node are designed separately according to a uniform constraint on the pruning error probability, which further reduce the number of scenarios that need to be calculated during the decoding procedure. In addition, by exploiting the properties of the joint weight distribution, a simplified calculation method of thresholds is proposed. Using this simplified calculation method, the number of scenarios that required to be calculated is reduced from $(d+1)(d+2)/2$ to $d+1$. 

%Simulation results show that the proposed method significantly reduces the decoding latency without tangibly degrading the error-correction performance.for insertion/deletion error channels
\end{abstract}

\IEEEpeerreviewmaketitle
\section{Introduction}
{D}eletion errors occur in the inter-vehicle communications\cite{VTP} and storage devices\cite{ISIT},  \cite{RACETCODE}, \cite{CSRM}, which result in clock synchronization errors. The loss of exact synchronization causes degradation of communication quality\cite{TITDI} and storage efficiency\cite{CSRM}. To tackle the error-correction problem in deletion error channels, a complementary skyrmion racetrack memory is proposed\cite{CSRM}, where the skyrmions are driven into two different nanotracks selectively with the aid of a voltage-controlled y-junction. However, the complementary structure introduces extra complexity and additional equipment overhead.
Besides, when the channels encounter synchronization errors, retransmissions or error-correction codes are often used to improve the transmission quality. However, retransmissions result in high delay and low transmission efficiency\cite{VTP}. Therefore, coding algorithms that can correct synchronization errors are urgently needed.

To deal with the synchronization errors, several error-correction codes have been proposed. The number-theoretic insertion/deletion-correcting codes are proposed to correct multiple insertion or deletion errors\cite{multiids}. Low-density parity-check codes (LDPC) connected with watermark codes are designed to correct insertion and deletion errors in differential pulse-position modulation systems\cite{CWGICC}, where synchronization errors are converted to substitution errors and corrected by LDPC codes. Synchronization error correction scheme using spatially coupled code was proposed in \cite{SGC}.

Polar codes \cite{POLAR} are capacity-achieving channel codes with a simple encoder and the successive cancellation (SC) decoder in discrete memoryless channels. With SC list decoding, polar codes outperform the existing turbo and LDPC codes\cite{SCL}, \cite{CRCSCL}. In addition, polar codes have been applied to deal with the error-correction problems in different types of scenarios, including the deletion channels \cite{vardy2017,TK2017}, \cite{TK2018}, \cite{TK2019}, \cite{ISIT2019}, insertion/deletion channels\cite{9568885} and control channels\cite{9756315,9309398}. The SC decoder proposed in \cite{ISIT2019} considers the insertion and deletion errors that occur with a certain probability, where the decoding complexity has not been reduced.
A list decoder based on SC decoding is proposed in \cite{vardy2017}, where all possible deletion patterns are considered in the list and exhaustively decoded by the SC decoder. Since the positions of deletion errors are unknown, there are $\bigl(\begin{smallmatrix} N  \\ d  \end{smallmatrix}\bigr)$ deletion patterns that need to be considered during the decoding procedure under the $d$-deletion channel, which causes high decoding complexity. Let $\cal O$ be the order of magnitude. The decoding complexity of the SC decoder proposed in \cite{vardy2017} is $\cal {O}$$(N^{d+1}\text{log}N)$. To reduce the decoding complexity, a modified SC decoding algorithm jointly considers multiple scenarios when decoding each node, and the complexity is reduced to $\cal O$$(d^{2}N\text{log}N)$\cite{TK2017}, \cite{TK2018}. The scenario-simplified successive cancellation (SSSC) decoding of polar codes further reduces the decoding complexity by pruning some low-probability scenarios\cite{TK2019}. However, the occurrence probability distribution of the scenarios varies among different nodes, which has not been considered by the greedy pruning in the SSSC decoding procedure. Due to this defect, the greedy pruning in the SSSC decoder results in a very large pruning error rate on some nodes, which leads to severe degradation of decoding performance. 

To address these problems, a new pruning strategy is proposed in this paper. The main contributions are summarized as follows
\begin{itemize}
  \item Considering that the occurrence probability distribution of the scenarios varies among different nodes, the pruning threshold for each node is separately designed, which maximizes the number of scenarios being pruned. Different from the greedy pruning in the SSSC decoder, the proposed pruning method ensures that the pruning error probability of each node is uniformly limited, which avoids the degradation of decoding performance caused by some nodes with extremely large pruning error rates.
  \item Further, a simplified calculation method is designed to obtain the pruning thresholds, where the occurrence probability distribution of scenarios is modeled by the product of two random variables subjecting to the hypergeometric distribution. Taking advantage of the nature of hypergeometric distribution, the $d$ peak values of the scenario occurrence probabilities are obtained directly without calculating all the scenario occurrence probabilities. Then the peak values are used to calculate the pruning thresholds, which avoids the probabilities calculation of all the $(d+1)(d+2)/2$ scenarios, thereby reducing the computational complexity from $\cal O$$\left( d^2 \right)$ to $\cal O$$\left( d \right)$.
  
\end{itemize} 

\section{Preliminaries}\label{sec002}

\subsection{ Polar Codes }

A polar code of length $N={2^n}$ is generated by
$\begin{matrix}  x_1^N = u_1^N{ {F} ^{ \otimes n}}, \end{matrix}$
where $F$=$\left[ {\begin{array}{*{20}{c}}
1&0\\
1&1
\end{array}} \right]$. ${u_1^N}=\left( u_1, u_2, \ldots , u_N \right)$ and ${x_1^N}=\left( x_1, x_2, \ldots , x_N \right)$ denote the codeword and coded sequence. ${ \otimes n}$ is the $n$-th Kronecker product. Through the channel polarization, $N$ polarized bit-channels have different reliabilities. The bit-channels with relatively lower reliability are selected to carry the information bits. ${\cal F}$ and ${\cal F}^c$ are defined as the set of frozen and infirmation bits, respectively. ${\cal H}\left( {p,m,n} \right)$ denotes the hypergeometric distribution with expectation $p\frac{m}{n}$ and variance $p\frac{m}{n}\left( {1 - \frac{m}{n}} \right)\frac{{n - p}}{{n - 1}}$.

\subsection{Noisy D-Deletion Channel Model}

\begin{figure}[H]
\begin{center}
\centering
\includegraphics[scale=0.39613]{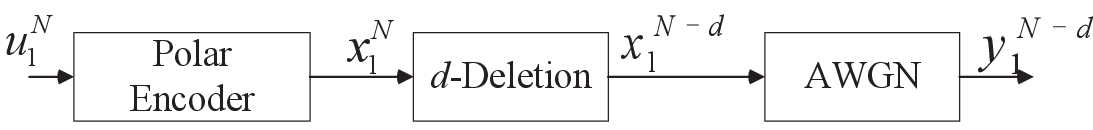} % 0.39786
\caption{Noisy $d$-deletion channel in cascade with the AWGN channel.}
\label{Fig001}\end{center}
\end{figure}

The $d$-deletion channel model is shown in Fig. \ref{Fig001}, which is an deletion model in cascade with the AWGN channel. This model is given by \cite{TK2017}\cite{TK2019}. After encoding, the codewords $x^N_1$ are transmitted through the $d$-deletion channel with added gauss noise, where the channel deletes $d$ symbols and outputs $y^{N-d}_1$. The receiver is unaware of the positions of the deleted symbols, and the number of deletion error patterns is $\bigl(\begin{smallmatrix} N  \\ d  \end{smallmatrix}\bigr)$. We adopt the assumption that used in \cite{TK2019}, where the probability of each symbol being deleted is the same. Therefore, each deletion pattern has the same occurrence probability.

\subsection{SC Decoding Method in D-Deletion Channel}
\begin{figure}[H]
\begin{center}
\centering
\includegraphics[scale=0.6]{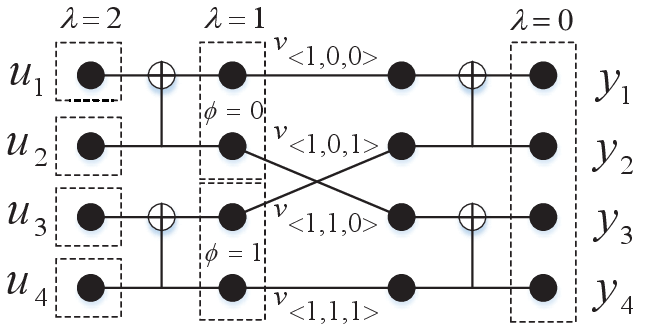} % 0.39786
\caption{Factor graph of polar codes with $N=4$.}
\label{fig1}\end{center}
\end{figure}

After receiving $y^{N-d}_1$, the SC decoder of polar codes estimates $u_i$ in a sequential manner, where the transition probabilities of sub-channels are recursively calculated by

\begin{equation}
\begin{split}
\qquad W_{2N}^{\left( {2i - 1} \right)}&\left( {y_1^{2N - d},u_1^{2i - 2}|{u_{2i - 1}}} \right)\\
& = \frac{1}{2}\sum\limits_{t = 0}^d {\frac{1}{{\left( {\begin{array}{*{20}{c}}
{2N}\\
d
\end{array}} \right)}}\left( {\begin{array}{*{20}{c}}
N\\
t
\end{array}} \right)} \left( {\begin{array}{*{20}{c}}
N\\
{d - t}
\end{array}} \right)\\
 & \times \sum\limits_{{u_{2i}}} {W_N^{\left( i \right)}\left( {y_1^{N - t},u_{1,e}^{2i - 2} \oplus u_{1,o}^{2i - 2}|{u_{2i - 1}} \oplus {u_{2i}}} \right)} \\
 & \times W_N^{\left( i \right)}\left( {y_{N - t + 1}^{2N - d},u_{1,e}^{2i - 2}| {u_{2i}}} \right),
\end{split}
\end{equation}
and
\begin{equation}
\begin{split}
W_{2N}^{\left( {2i} \right)}&\left( {y_1^{2N - d},u_1^{2i - 1}|{u_{2i}}} \right)\\
 & = \frac{1}{2}\sum\limits_{t = 0}^d {\frac{1}{{\left( {\begin{array}{*{20}{c}}
{2N}\\
d
\end{array}} \right)}}\left( {\begin{array}{*{20}{c}}
N\\
t
\end{array}} \right)} \left( {\begin{array}{*{20}{c}}
N\\
{d - t}
\end{array}} \right)\\
 & \times W_N^{\left( i \right)}\left( {y_1^{N - t},u_{1,e}^{2i - 2} \oplus u_{1,o}^{2i - 2}|{u_{2i - 1}} \oplus {u_{2i}}} \right)\\
 & \times W_N^{\left( i \right)}\left( {y_{N - t + 1}^{2N - d},u_{1,e}^{2i - 2}|{u_{2i}}} \right),
\end{split}
\end{equation}

For notation convenience, we temporarily discuss the notations on the factor graph of polar codes in memoryless channel. Fig. \ref{fig1} shows the factor graph of the polar code with $N=4$, where the layers from left to right are labeled by $\lambda  = 2,\lambda  = 1,\lambda  = 0$. Each layer is divided into ${2^\lambda }$ groups, where each group contains ${2^{n - \lambda }}$ elements. The label of each group is denoted by $\phi $, and the index of the element is $\beta$. The location of each node is determined by a set of parameters $(\lambda,\phi ,\beta)$. Let ${v_{ < \lambda ,\phi ,\beta  > }}$ denote the index of each node in the factor graph. When decoding node $v$, the coded bits related to this node can be represented by $x_{\beta {2^\lambda } + 1}^{\left( {\beta  + 1} \right){2^\lambda }}$, which means that the corresponding channel outputs of $x_{\beta {2^\lambda } + 1}^{\left( {\beta  + 1} \right){2^\lambda }}$ are involved in the calculation procedure of the transition probability at node ${v_{ < \lambda ,\phi ,\beta  > }}$. According to the relative positions of the sequence $x_{\beta {2^\lambda } + 1}^{\left( {\beta  + 1} \right){2^\lambda }}$, the coded bits are decomposed into three parts: $x_1^{\beta {2^\lambda }}$, $x_{\beta {2^\lambda } + 1}^{\left( {\beta  + 1} \right){2^\lambda }}$ and $x_{\left( {\beta  + 1} \right){2^\lambda } + 1}^N$. Let ${N_1},{N_2},{N_3}$ be the lengths of the three parts, where ${N_1} = \beta {2^\lambda },{N_2} = {2^\lambda },{N_3} = {2^n} - \left( {\beta  + 1} \right){2^\lambda }$. The empty set is denoted by ${\O}$.
For example, the coded bits corresponding to the decoding procedure of node $v_{ < 1, 0, 0> }$ is $x^2_1$, and the whole coded bit sequence is decomposed into ${\O}$, $x^2_1$ and $x^4_3$. ${N_1}=0,{N_2}=2,{N_3}=2$. While the coded bits corresponding to the decoding procedure of node $v_{ < 1, 0, 1> }$ is $x^4_3$, and the whole sequence is decomposed into $x^2_1$, $x^4_3$ and ${\O}$. ${N_1}=2,{N_2}=2,{N_3}=0$.

In the deletion channel, the received symbols corresponding to $x_{\beta {2^\lambda } + 1}^{\left( {\beta  + 1} \right){2^\lambda }}$ vary with the number of deletion errors. Define ${d_1},{d_2},{d_3}$ as the number of deletion errors in $x_1^{\beta {2^\lambda }}$, $x_{\beta {2^\lambda } + 1}^{\left( {\beta  + 1} \right){2^\lambda }}$ and $x_{\left( {\beta  + 1} \right){2^\lambda } + 1}^N$, respectively. With fixed $d_1, d_2$, the received symbols corresponding to $x_{\beta {2^\lambda } + 1}^{\left( {\beta  + 1} \right){2^\lambda }}$ are $y_{\beta {2^\lambda } + 1 - {d_1}}^{\left( {\beta  + 1} \right){2^\lambda }-{d_1}-{d_2}}$. A node with a specific deletion pattern $< {d_1},{d_2},{d_3} >$ is named as a \emph{scenario}, which can be represented by $v_{ < \lambda ,\phi ,\beta  > }^{ < {d_1},{d_2} > }$. In other words, the combination of the specific number $< {d_1},{d_2},{d_3} >$ of deletion errors for node $v$ is called a scenario.

For each node ${v_{ < \lambda ,\phi ,\beta  > }}$, different scenarios correspond to different numbers of deletion errors in the segment $x_1^{\beta {2^\lambda }}$, $x_{\beta {2^\lambda } + 1}^{\left( {\beta  + 1} \right){2^\lambda }}$ and $x_{\left( {\beta  + 1} \right){2^\lambda } + 1}^N$. The SC decoder needs to calculate the transition probability separately for each scenario. Therefore, the number of scenarios determines the decoding complexity of the SC decoder in the deletion channels. Besides, the calculation of transition probability is the main part of the decoding complexity in the SC decoding procedure. The traditional SC decoding in AWGN channel can be regarded as a special case of the noisy $d$-deletion channel with $d=0$, which means that only one transition probability needs to be calculated on each node of the factor graph. The number of transition probabilities that needs to be calculated by the traditional SC decoder is $N(1+\text{log}(N))$, and the decoding complexity is $N\text{log}(N)$. For the $d-$deletion channel, the transition probability needs to be calculated for each scenario on every node. Since there are at most $(d+2)(d+1)/2$ scenarios in each node, the decoding complexity of the SC decoding method in deletion channel is $\cal O$$\left( {{d^2}N\text{log} N} \right)$. The decoding complexity is mainly determined by the transition probability calculation operations and the number of scenarios. For each scenario, the decoder needs to perform one transition probability calculation operation, which means that each scenario causes large decoding complexity. Therefore, reducing the number of patterns is important for reducing the overall decoding complexity.

Moreover, the occurrence probability of each scenario is called the \emph{joint weight}\cite{TK2019}, which is defined as

\begin{equation}
J \left( v_{ < \lambda ,\phi ,\beta  > }^{ < {d_1},{d_2} > } \right)= \frac{{\left( {_{{d_1}}^{{N_1}}} \right)\left( {_{{d_2}}^{{N_2}}} \right)\left( {_{{d_3}}^{{N_3}}} \right)}}{{\left( {_d^N} \right)}}.
\label{eq6}\end{equation}

According to (\ref{eq6}), the joint weight is determined by the location of the nodes and the code length $N$, which can be calculated once the code length and the number of deletion errors are known.

\section{SC Decoding Based on Pruning Strategy }\label{sec003}

Reducing the number of scenarios that involved in the decoding procedure is important for the practical implication of polar codes in deletion channels. According to (\ref{eq6}), the joint weights of scenarios are different. However, the decoder needs to calculate a transition probability for each scenario, which means that each scenario leads to a similar procedure of calculating the transition probability regardless of its probability of occurrence. Therefore, the scenarios with small joint weight can be pruned to reduce the decoding complexity without significant performance loss.

The SSSC method \cite{TK2019} uses the same value as the threshold to prune the scenarios. This unified pruning threshold ignores the differences of the joint weight distribution among different nodes, which might lead to extremely large node pruning error probability on some nodes. The extremely large node pruning error probability on some nodes causes significant decoding performance loss. To deal with these problems, we design a novel pruning strategy to avoid the occurrence of nodes with extremely high pruning error probability, where different pruning thresholds are separately designed for different nodes.

\subsection{Pruning Strategy Based on Pruning Error Bounds}\label{sec3}

In order to avoid the occurrence of nodes with large pruning error probability and achieve accurate pruning, we first analyze the error probability caused by pruning on each node. Let ${\cal P} = \left\{ {{S_1},{S_2}, \cdots ,{S_m}} \right\}$ denote the pruning pattern. The pruning error probability of each node is obtained by

\begin{equation}
Pe = \sum\limits_{i = 1,{{{S_i}}} \in {\cal P}}^m {{J_{{S_i}}}},
\label{eq7}\end{equation}
where $J_{S_i}$ denotes the joint weight of the scenario $S_i$.

\begin{figure}[h]
\begin{center}
\hspace{-0.19613cm}
\centering
\includegraphics[scale=0.51316]{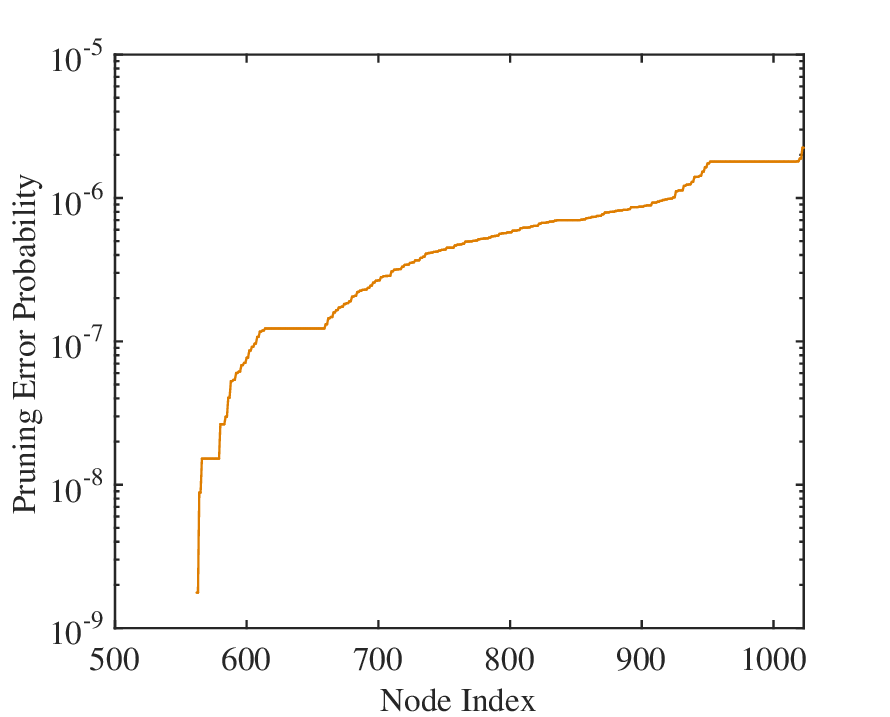} % 0.39786
\caption{Pruning Error Probability of the SSSC pruning. ($d=5$.)}
\label{F001}\end{center}
\end{figure}

In the SSSC decoder, the pruning threshold $\tau_1$ is a unified limit on the joint weight of scenarios for all nodes. However, this unified threshold is unreasonable, because it ignores the difference of joint weight distribution and might lead to a large pruning error probability on some nodes with relatively uniform joint weight distribution. For example, for the (512, 256) polar codes, the node error probabilities caused by pruning operations are shown in Fig. \ref{F001}, where the pruning threshold is set to $\tau_1=10^{-6}$ in the SSSC decoding procedure. It shows that the pruning error probability of different nodes is on different orders of magnitude, where the smallest pruning error probability is lower than  $10^{-8}$, while the larger one is higher than $10^{-6}$. If a uniform pruning threshold $\tau_1$ is adopted, the pruning error probability of some nodes will be very small, while some will be very large. For the nodes with a small pruning error probability, more scenarios can be pruned without significant degradation of the decoding performance. For the nodes with a large pruning error probability, the unified pruning threshold results in a significant decoding performance loss and high error floors. Therefore, if the pruning error probability of different nodes can be uniformly limited, the decoding performance and complexity can be better balanced, and the pruning can be more effective and sufficient.

Inspired by this, we propose a pruning strategy with performance constraints (PSPC), where the pruning error probability are limited by a uniform bound $Pe_{\text{bound}}$. Let $\theta$ denote the number of scenarios in node $v$. Let $\left\{ {{J_{{S_i}}}|i = 1, \cdots ,\theta} \right\}$ denote the set of joint weights that ranked in an ascending order. First, we determine the maximum number $k$ of pruned scenarios by

\begin{equation}
\begin{array}{*{20}{c}}
k = {\mathop {\text{max} }\limits_{m } \left\{ m | \sum\limits_{i = 1}^m {{J_{{S_i}}}}\leq Pe_{\text{bound}} \right\}}.
\end{array}
\label{eq8}\end{equation}

The pruning threshold $\tau_2$ of each node is obtained by

\begin{equation}
{\tau_2} = J_{{S_k}}.
\label{eq901}\end{equation}

In particular, when $k=0$, we have $\tau_2=0$, which means that the scenario with the smallest joint weight has already exceeded the upper limit of pruning errors and no scenario will be pruned. After obtaining the threshold by (\ref{eq901}), the scenarios whose joint weight is less than or equal to $\tau_2$ will be pruned.

Different from the uniform pruning threshold used in the SSSC decoding, the PSPC pruning method adopts a uniform limit on the sum of the joint weights of the pruned scenarios, which avoids the nodes with extremely high pruning error probability. The proposed pruning thresholds are independently determined for each node according to their joint weight distributions. At the given limit of the pruning error probability, the proposed method prunes as many scenarios as possible, which significantly reduces the decoding complexity. 

\subsection{Simplified Calculation Method of Pruning Thresholds}\label{sec5}

According to (\ref{eq8}), the joint weights of all scenarios are needed to be calculated to obtain the pruning threshold. The maximum number of the scenarios is $\left( {d + 1} \right)\left( {d + 2} \right)/2$ for each node, which causes high computational complexity. To deal with this problem, we analyze the property of joint weight distribution and propose a simplified calculation method to determine the pruning threshold.
According to the definition of joint weight and its probability distribution, for each node ${v_{\left\langle {\lambda ,\phi ,\beta } \right\rangle }}$, the joint weight of scenario $S=v_{ < \lambda ,\phi ,\beta  > }^{ < {d_1},{d_2} > }$ can be represented by the product of two parts as follows,

\begin{equation}
J_{S} = \frac{{\left( {_{{d_1}}^{{N_1}}} \right)\left( {_{{d_2}}^{{N_2}}} \right)}}{{\left( {_{{d_1} + {d_2}}^{{N_1} + {N_2}}} \right)}} \times \frac{{\left( {_{{d_1} + {d_2}}^{{N_1} + {N_2}}} \right)\left( {_{{d_3}}^{{N_3}}} \right)}}{\left( {_{{d_1} + {d_2} + {d_3}}^{{N_1} + {N_2} + {N_3}}} \right)}.
\label{eq10}\end{equation}

\begin{table}[h]
\caption{Joint Weight of Scenarios.}
\label{tab:1}       % Give a unique label
\centering\begin{tabular}{  p{1cm}<{\centering} p{1cm}<{\centering} p{1cm}<{\centering}  p{1cm}<{\centering}  p{2cm}<{\centering} }
\hline\noalign{\smallskip}
Index & ${d_3}$ & ${d_1}$ & ${d_2}$ & Joint Weight \\
\noalign{\smallskip}\hline\noalign{\smallskip}
1&0 & 0 & 3 & 0 \\
2&0 & 1 & 2 & 0.0071 \\
3&0 & 2 & 1 & 0.0214 \\
4&0 & 3 & 0 & 0.0071 \\
5&1 & 0 & 2 & 0.0178 \\
6&1 & 1 & 1 & 0.1428 \\
7&1 & 2 & 0 & 0.1071 \\
8&2 & 0 & 1 & 0.1607 \\
9&2 & 1 & 0 & 0.3214 \\
10&3 & 0 & 0 & 0.2142 \\
\noalign{\smallskip}\hline
\end{tabular}
\end{table}

\begin{figure}[h]
\begin{center}
\hspace{-0.39613cm}
\centering
\includegraphics[scale=0.51316]{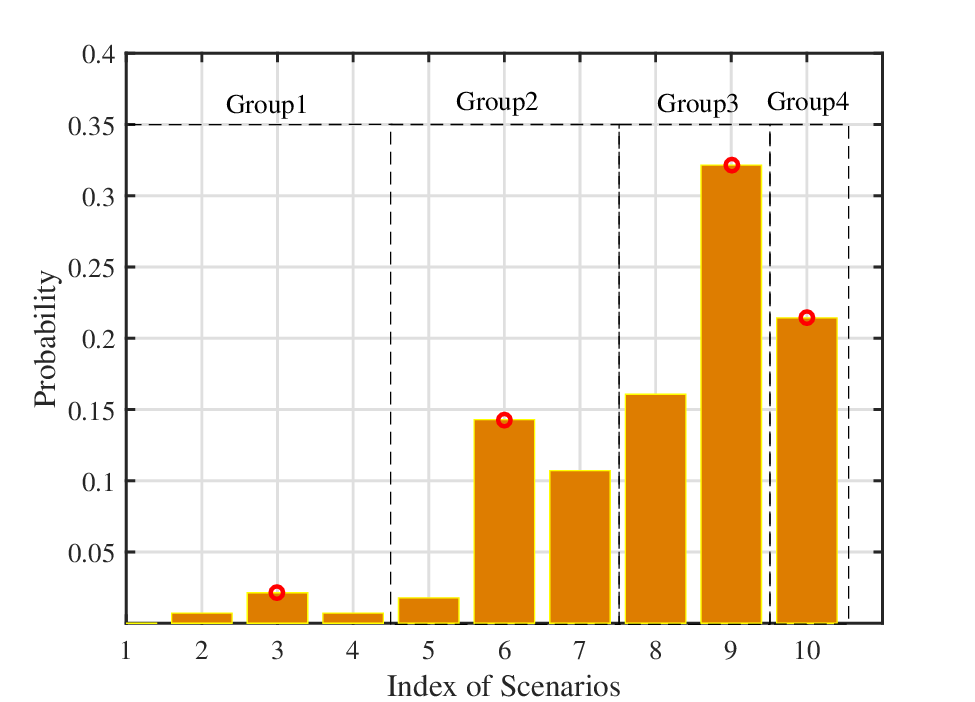} % 0.39786
\caption{Probability Distribution of Scenarios in $v_{ < 1 ,1 ,2  > }$. ($N=16$, $d=3$)}
\label{fig2}\end{center}
\end{figure}

Let $P_1$ and $P_2$ denote the first part $\frac{{\left( {_{{d_1}}^{{N_1}}} \right)\left( {_{{d_2}}^{{N_2}}} \right)}}{{\left( {_{{d_1} + {d_2}}^{{N_1} + {N_2}}} \right)}}$ and the second part $\frac{{\left( {_{{d_1} + {d_2}}^{{N_1} + {N_2}}} \right)\left( {_{{d_3}}^{{N_3}}} \right)}}{{\left( {_d^N} \right)}}$, respectively. The first part $P_1$ is the probability density function (PDF) of a random variable which is subject to the hypergeometric distribution ${\cal H}\left( {{d_1} + {d_2},{N_1},{N_1} + {N_2}} \right)$, and the second part $P_2$ denotes the PDF of the variable that subjects to ${\cal H}\left( {d,{N_3},N} \right)$. Since ${d_3}$ is only related to the second part $P_2$, the scenarios with the same ${d_3}$ are listed into a group. For example, Fig. \ref{fig2} and Table \ref{tab:1} show the joint weight distribution of scenarios in node $v_{ < 1 ,1 ,2  > }$ which are grouped into 4 subgroups in Table \ref{tab-2}.

\begin{table}[h]
\caption{Group Indexes and Peak Values of Different Groups.}
\label{tab-2}       % Give a unique label
\centering\begin{tabular}{  p{1.5cm}<{\centering}  p{0.6cm}<{\centering}  p{1.5cm}<{\centering}  p{1.5cm}<{\centering} p{1.5cm}<{\centering} }
\hline\noalign{\smallskip}
Group Index & ${d_3}$ & ${d_1}$ at peak & ${d_2}$ at peak & Peak Value \\
\noalign{\smallskip}\hline\noalign{\smallskip}
1 & 0 & 2 & 1 & 0.0214 \\
2 & 1 & 1 & 1 & 0.1428 \\
3 & 2 & 1 & 0 & 0.3214 \\
4 & 3 & 0 & 0 & 0.2142 \\
\noalign{\smallskip}\hline
\end{tabular}
\end{table}

In each node, $N_1$, $N_2$, $N_3$ and $d$ are fixed, and the joint weight varies with the combination of $d_1, d_2, d_3$. For a given $d_3$, $P_2$ is a constant. Therefore, in each subgroup, the joint weight of scenarios varies with $d_1$ and $d_2$, where the range of variable $d_1$ is ${d_1} = \left\{ {0,1, \cdots ,d - {d_3}} \right\}$ and ${d_2} = d - {d_1} - {d_3}$.

Besides, for a random variable $X$$\sim$${\cal H}\left( {p, m, n} \right)$, the PDF is
\begin{equation}
\text{P} \left( X = k \right) = \frac{{\left( {_{{{k}}}^{{m}}} \right)\left( {_{{{p-k}}}^{{n-m}}} \right)}}{{\left( {_{p}^{n}} \right)}},
\label{eq630001}\end{equation}

The probability increases first and then decreases with the increase of $k$, while it reaches the peak value at ${{k}} = \left\lceil {\frac{{{m} p }}{{n}}} \right\rceil $. Therefore, the peak value of $P_1$ at given $d_3$ is

\begin{equation}
\text{max} \left\{ {{P_1}} \right\} = \frac{{\left( {_{{d_{1m}}}^{{N_1}}} \right)\left( {_{{d_{2m}}}^{{N_2}}} \right)}}{{\left( {_{d_1+d_2}^{N_1+N_2}} \right)}},
\label{eq631}\end{equation}
where ${d_{1m}} = \left\lceil {\frac{{{N_1}\left( {{d} - {d_3}} \right)}}{{{N_1} + {N_2}}}} \right\rceil $ and ${d_{2m}} = d - \left\lceil {\frac{{{N_1}\left( {{d} - {d_3}} \right)}}{{{N_1} + {N_2}}}} \right\rceil  - {d_3}$.

Then we have the following remark.
\begin{remark} For each ${d_3}$, the peak value $\gamma$ of a subgroup is
\begin{equation}
\gamma \left( {{d_3}} \right) = \frac{{\left( {_{{d_{1m}}}^{{N_1}}} \right)\left( {_{{d_{2m}}}^{{N_2}}} \right)\left( {_{{d_3}}^{{N_3}}} \right)}}{{\left( {_d^N} \right)}}.
\label{eq11}\end{equation}
\label{Remark1}\end{remark}

Rank the peak values in an ascending order and obtain the set $\Omega $, where $\Omega  = \left\{ {\delta \left( i \right)|\delta \left( i \right) \le \delta \left( {i + 1} \right),i = 1, \cdots ,d} \right\}$ and ${\delta} \in \left\{ {\gamma \left( {d_3} \right)|{d_3} = 0,1, \cdots ,d} \right\}$. From Fig. \ref{fig2}, the peak value of each group can approximately reflect the probability level of this subgroup. Therefore, we adopt the peak values of subgroups to approximate the probability distribution of all the $(d+2)(d+1)/2$ scenarios in each node. Then, we adopt the peak values of subgroups to determine the pruning thresholds and propose a simplified pruning strategy with performance constraints (SPSPC). Let ${\eta _c} = \sum\limits_{i = 1}^c {{\delta _i}} $ denote the partial sum of elements in set $\Omega $. Assuming that all scenarios with a probability smaller than ${\delta _t}$ are pruned, the relationship between the pruning error probability $Pe_{\text{bound}}$ and the accumulation of peak values $\eta _t$ can be approximated by,

\begin{equation}
\frac{{{\eta _t}}}{{{\eta _{d + 1}}}}=\left( {\sum\limits_{i = 1}^t {{\delta _i}} } \right)/\left( {\sum\limits_{i = 1}^{d + 1} {{\delta _i}} } \right) \cong Pe_{\text{bound}}.
\end{equation}

Then the simplified calculation method of thresholds is defined by proposition 1.

\begin{proposition}
The pruning threshold is defined as follows,

\begin{equation}
\tau_2  = \delta \left( k \right),k =\mathop {\max }\limits_{t \in \{1,\cdots,d+1\}} \left\{ t | {\eta _t} \le Pe_{\text{bound}}  \cdot {\eta _{d + 1}} \right\} ,
\label{eq13}\end{equation}

Specifically, ${\eta _1} > Pe_{\text{bound}}  \cdot {\eta _{d + 1}} $ denotes that all peak values are larger than $ Pe_{\text{bound}}  \cdot {\eta _{d + 1}}$, which means that the performance loss caused by pruning is small. Then the pruning threshold is set to

\begin{equation}
\tau_2  = Pe_{\text{bound}}  \cdot {\eta _{d + 1}} .
\label{eq14}\end{equation}
\end{proposition}

Scenario $S$ with weight $J_S\leq\tau_2$ will be pruned and the calculation of its transition probability will be omitted during the decoding procedure. Since only the peak values are considered, the number of joint weights that need to be calculated is reduced from $(d+2)(d+1)/2$ to $(d+1)$.

\subsection{Storage Complexity Analysis}

The thresholds for the proposed pruning strategy are determined by the joint weight distribution and the constraint coefficient $Pe_{\text{bound}}$. Therefore, when the code length and the number of deletion errors are determined, the thresholds can be calculated in an off-line manner. Since the joint weight distributions of scenarios are different among different nodes, the thresholds $\tau_2$ of different nodes are also different, which need to be calculated separately. According to the definition of the joint weight (i.e., Eq. (\ref{eq6})), the nodes in different groups with the same $\beta$ and the same layer index $\lambda$ have the same joint weight distribution of scenarios. Therefore, the same pruning threshold can be reused among the nodes with the same $\beta$ index in the same layer. Observing the structure of the factor graph, there are $2^{n-\lambda}$ nodes in each group of the $\lambda$-th layer, which means that the number of thresholds that need to be calculated in the $\lambda$-th layer is $2^{n-\lambda}$. Therefore, the number of pruning thresholds that needed to be calculated and stored is

\begin{equation}
\begin{array}{*{20}{c}}
\sum\limits_{\lambda = 1}^{n-1} {{2^{{n-\lambda}}}} = N-2.
\end{array}
\label{eqs002}\end{equation}
By taking advantage of the periodicity of the joint weight distribution, the proposed pruning threshold only requires linear storage complexity.

\section{Conclusion}\label{sec005}

This paper proposes a novel pruning strategy to reduce the decoding complexity of polar codes in deletion channels, where a uniform constraint on the sum of the joint weight of the pruned scenarios is used to determine the pruning thresholds for effective pruning. Moreover, the proposed pruning method avoids the nodes with extremely high decoding error probability, which leads to better decoding performance. Moreover, to simplify the calculation of pruning thresholds, a simplified SPSPC method is designed, which takes advantage of the properties of the weight distribution and reduces the computational complexity from $\cal O$$(d^2)$ to $\cal O$$(d)$. 

\begin{table}
\caption{The Number of Scenarios Involved in Decoding. }
\label{tab-3}       % Give a unique label
\centering\begin{tabular}{  p{1cm}<{\centering}  p{1.6cm}<{\centering}  p{1.3cm}<{\centering} p{1.3cm}<{\centering} p{1.3cm}<{\centering}}
\hline\noalign{\smallskip}
$N$ & SC\cite{TK2017} & SSSC\cite{TK2019} & SC-SPSPC & Reduction\\
\noalign{\smallskip}\hline\noalign{\smallskip}
512 & 82944 & 63074 & 54514 & 8560 \\
1024 & 1003904 & 572500 & 537052 & 35448 \\
2048 & 2834432 & 1203772 & 1180508 & 23264 \\
\noalign{\smallskip}\hline
\end{tabular}
\end{table}

\bibliographystyle{IEEEtran}
\bibliography{mybib}

\end{document}